\documentclass[
]{ceurart}

\usepackage{listings}
\usepackage{todonotes}
\usepackage{booktabs}
\usepackage{comment}
\usepackage[super]{nth}

\makeatletter
\newcommand*{\rom}[1]{\expandafter\@slowromancap\romannumeral #1@}
\makeatother

\def\footnoterule{\relax%
  \kern-5pt
  \hbox to \columnwidth{\hfill\vrule width 0.5\columnwidth height 0.4pt\hfill}
  \kern4.6pt}
\makeatother

\makeatletter
\newcommand{\linebreakand}{%
  \end{@IEEEauthorhalign}
  \hfill\mbox{}\par
  \mbox{}\hfill\begin{@IEEEauthorhalign}
}
\makeatother 

\begin{document}

\copyrightyear{2024}
\copyrightclause{Copyright for this paper by its authors.
  Use permitted under Creative Commons License Attribution 4.0
  International (CC BY 4.0).}

\conference{ }
%\conference{AICS'24: 32nd Irish Conference on Artificial Intelligence and Cognitive Science,
%December 09--10, 2024, Dublin, Ireland}

\title{Back-filling Missing Data When Predicting Domestic Electricity Consumption From Smart Meter Data}

\author[1]{Xianjuan Chen}

\address[1]{School of Computing, Dublin City University, Glasnevin, Dublin 9, Ireland.}
\address[2]{Insight Research Ireland Centre for Data Analytics, Dublin City University, Glasnevin, Dublin 9, Ireland.}

\author[1]{Shuxiang Cai}

\author[2]{Alan F. Smeaton}[%
orcid=0000-0003-1028-8389,
email=Alan.Smeaton@DCU.ie,
]
\cormark[1]

\cortext[1]{Corresponding author.}

\maketitle

%%%%%%%%%%%%%%%%%%%%%%%%%%%%%%%%%%%%%%%%%%%%%%%%%%%%%%%%%%%%%
\begin{abstract} 
This study uses data from domestic electricity smart meters to estimate annual electricity bills for a whole year. We develop a method for back-filling data smart meter for up to six missing months for users who have less than one year of smart meter data, ensuring reliable estimates of annual consumption. We identify five distinct electricity consumption user profiles for homes based on day, night, and peak usage patterns, highlighting the economic advantages of Time-of-Use (ToU) tariffs over fixed tariffs for most users, especially those with higher nighttime consumption. Ultimately, the results of this study empowers consumers to manage their energy use effectively and to make informed choices regarding electricity tariff plans.
\end{abstract}
%-----------------------------------------------------------


\section{Introduction}

Since the initiation of the National Smart Metering Programme in late 2019, Ireland has been making progress towards installing over 2 million smart meters by early 2025~\cite{ESBN.2023}. These smart meters empower customers to better manage their electricity consumption by using recorded usage information. However, the abundance of suppliers offering slightly varied rates makes selecting the most cost-effective time-of-use (ToU) tariff a challenging and confusing task for many households. Some people online~\cite{smartcost} are even worried that smart meter tariff plans will end up costing customers more. To address this complexity, recognising the potential and the limitations of historical data is essential, as data may be incomplete or biased due to factors like weather fluctuations and household behaviours. 

This study describes the development and implementation of a  model that identifies five typical electricity consumption profiles across a diverse consumer population. We then use these profiles to estimate consumption patterns for individual households who have up to six months of incomplete data, and recommend the cheapest ToU energy plan based on their consumption patterns.

\section{Background and Context}

\subsection{Smart Meters}
\label{subsec:SMs}
Smart meters are advanced electronic devices designed to measure both the amount of electricity exported to the grid and imported from the grid by a domestic or business customer. They offer consumers and energy providers detailed and up-to-date information on energy consumption compared to traditional meters by eliminating the necessity for approximate meter readings~\cite{esn,ACER.2023.Report6}.

The interval for metering and recording the consumption varies within the European Union (EU) from 15 minutes to 2 hours, depending on the country~\cite{ACER.2023.Report6} and in Ireland the set interval is 30 minutes, uploaded at the end of each day.
%As shown in Table~\ref{tab:frequency}, 
%
%\begin{table}[htb]
%    \centering
%    \caption{Smart meter measurement frequency across the EU countries}
%    \label{tab:frequency}
%    \begin{tabular}{p{1cm}|p{2cm}|p{1cm}|p{2cm}} 
%\toprule 
%Country & Measurement frequency&Country & Measurement frequency\\
%\midrule 
%AT & 15 min&IE & 30 min\\
%BE & N/A& IT & 15 min \\
%BG & N/A&  LT & 15 min\\ 
%CY & N/A& LU & 15 min\\
%CZ & 1 hour& LV & 1 hour\\
%DE & 15 min& MT & 1 hour\\
%DK & 15 min& NL & 15 min\\
%EE & 1 hour& NO& 1 hour\\
%ES & 1 hour& PL & 15 min \\
%FI & 1 hour& PT&15 min\\
%FR & 30 min& RO&15 min\\
%GR & N/A&SE&1 hour  \\
%HR & 1 hour&SI&15 min \\
%HU & 15 min&SK& 15 min\\
%\bottomrule
%    \end{tabular}
%\end{table}
%
%Through the use of smart meters, users can gain enhanced control over their energy consumption. 
Smart meters facilitate demand-side mechanisms such as ToU tariffs wherein consumers are charged different tariff rates based on various timeslots of the day. Higher rates are typically applied during peak demand, while lower rates are employed during off-peak periods. 
%With such features, 
Smart meters enable customers to 
%use electricity more efficiently by shifting energy usage from peak to off-peak times or 
select their cheapest tariff from an energy supplier and then enable energy suppliers to offer more comprehensive insights, including information on the total energy consumption and categorised appliance usage throughout the household. Overall, they contribute to reducing carbon emissions and fostering the development of a more sustainable energy system~\cite{Bager,Belton}.

\begin{figure*}[ht]
    \centering
    \includegraphics[width=0.8\linewidth]{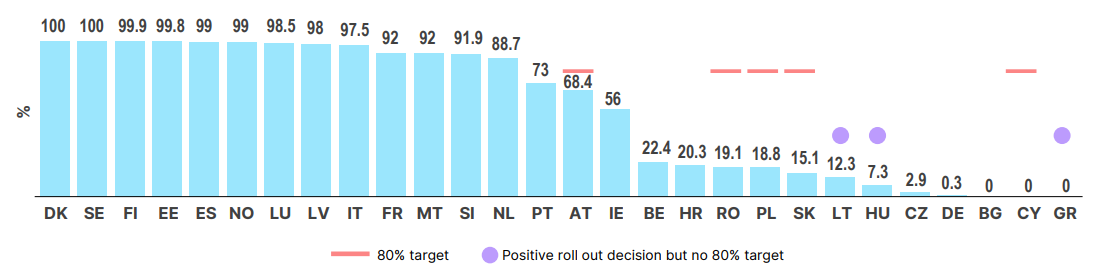}
    \caption{Installation rate for smart meters in the EU up to 2022~\cite{ACER.2023.Report6}.}
    \label{fig:Smart Meter Rollout Rate}
\end{figure*}

As illustrated in Figure~\ref{fig:Smart Meter Rollout Rate}, thirteen member states in the EU achieved a smart meter roll-out rate exceeding 80\% in 2022, whereas ten member states had rates below 20\%~\cite{ACER.2023.Report6}. In Ireland, ESB Networks is in the process of upgrading electricity meters nationwide and has announced the successful installation of 1.5 million smart meters in homes, farms and businesses across every county in Ireland in November 2023~\cite{ESBN.2023}. At the time of writing, the number of installations had risen to 1.8 million

\subsection{The Irish Energy Market}

The electricity sector in Ireland has undergone a paradigm shift, transitioning from a regulated monopoly to a deregulated competitive market~\cite{marketderegulated}.  Before the 1999 Act and the introduction of a supplier licensing regime, ESB held a monopoly on electricity retail supply but from February 2005 liberalisation measures empowered all Irish electricity consumers to choose their preferred suppliers~\cite{regulation}. This initially restricted ESB's ability to set retail prices independently, effectively diminishing its control over the market. As consumer switching activity reached a satisfactory level deemed by the Commission for Regulation of Utilities (CRU), these restrictions were progressively lifted for business and domestic consumers in 2010 and 2011. In 2012, ESB's retail division re-branded as Electric Ireland. By 2022, Electric Ireland dominated the domestic electricity market in Ireland with a 41\% share~\cite{ElectricitySupplyBoard}.

The deregulation of the Irish electricity market was positively received by both consumers and suppliers. This restructuring aimed to enhance consumer choice, lower retail prices, and introduce innovative products. Consumers became more engaged leading to a better understanding of energy costs and encouraging conservation efforts. The influx of competitive suppliers provided more choices and attractive pricing structures, allowing them to cater to unique customer needs and differentiate themselves by providing innovative services and benefits. The Consumer Association of Ireland echoed these sentiments, applauding the move as a ``win-win situation for suppliers and consumers"~\cite{deregulated}. 

The Commission for the Regulation of Utilities (CRU) functions as Ireland's independent regulator for energy and water, established in 1999 under the Irish Government's policy and framework~\cite{CRU}. The CRU's  oversees energy networks and promotes the use of renewable energy,  enforces quality standards, fosters competition, and provides consumers with information and tools for dispute resolution, and it regulates gas, petroleum, and electrical contractors.
It holds the authority to monitor and ensure that every licensed electricity supplier adheres to the terms specified in its supply license which cover aspects such as billing, disconnection, marketing, complaints handling, prepayment meters, and the treatment of vulnerable customers.

\label{sec:tariffs}

There is a list of all energy providers authorised by the CRU to furnish electricity and gas within the retail energy market in Ireland~\cite{supplier}. Presently, 13 energy suppliers serve the Irish Electricity market, catering to a population of 1.84 million households~\cite{census}. These are Arden Energy, Bord Gáis Energy, Ecopower, Energia, Electric Ireland, Flogas, Glow Power, Pinergy, PrePayPower, Yuno Energy Ltd, SSE Airtricity, Community Power, and Water Power. In total, they provide more than 60 tariffs for customers with smart meters as well as different fixed standing charge amounts depending on whether the household location is urban or rural.

\subsection{Electricity Consumption Profiles}
\label{subsub:profiles}

Many studies have  investigated domestic electricity consumption profiles across the world and one line of research has explored the relationship between the fundamental characteristics of households and their electricity usage patterns. Studies reported in~\cite{ANDERSEN2021105341, ANDREASGUNKEL2023108852} utilising electricity consumption data from Danish households revealed that heat pumps, electric vehicles (EVs), and dwelling types exert significant influence on consumption levels, while socio-economic factors like occupancy, dwelling area, and income have minimal impact. Additionally, Munkhammar~\cite{MUNKHAMMAR2015439} discovered that houses equipped with EV charging predominantly see an increase in electricity usage during the evening while research in Germany~\cite{WITTENBERG2016199} and Australia~\cite{DENG2017313} demonstrated that households with photovoltaic (PV) systems tend to have increased electricity consumption.

The analysis of consumption data has included statistical profile analysis and clustering for profile extraction~\cite{ZHANG2021116452}. Multiple studies~\cite{KANG2023112753,MICHALAKOPOULOS2024122943,RAJABI2020109628, CZETANY2021111376} have evaluated various clustering algorithms and found K-means has a consistent ability to yield superior results in this context. These studies have also emphasised the importance of determining the appropriate number of clusters and we build on the lessons and conclusions from this previous work in our use of K-means to cluster our used into 5 profiles.

\section{Experimental Setup}

\subsection{Data Collection}
\label{sec:data collection}
The data collection phase for this study was conducted from January 2024 to June 2024. During this period, two  categories of data were gathered: electricity tariff plans from energy suppliers and smart meter data from households.

The tariff plan data were obtained directly from the official websites of 10 of the 13 energy suppliers. We excluded 3 official suppliers because their market penetration was  small.  As of June 2024, irrespective of whether the household location is urban or rural, 58 unique electricity tariff plans were then available, divided into either fixed-rate or time-of-use (ToU) tariff plans. Specifically, 17 fixed-rate plans provide a constant price for electricity regardless of the time of day, while 41 ToU plans vary the price based on the time of consumption and this includes both day/night and smart tariffs.  The tariff plan details on the websites are updated periodically, influenced by  the energy market thus checks on the tariffs are performed every two weeks by manually visiting the official websites.

ESB Networks who provide and maintain the electricity generation and distribution, allows customers to download their own smart meter data. Customers  register on the ESB Networks website using their MPRN number and download a CSV file that includes all their electricity imports plus exports (if they micro-generate) from the installation date of their smart meter up to the previous day. This file is referred to as an HDF file, with a sample shown in the table.~\ref{tab:HDF}. 

\begin{table}[htb]
    \centering
    \caption{Excerpt of HDF file showing import and export over a 90 minute period.}
    \label{tab:HDF}
    \begin{tabular}{l|l|l|l}
\toprule 
MPRN & Value & Read Type & Read Date and Time\\
\midrule 
10000000000 & 0.007 & Export (kW) & 30-04-2024 12:30 \\
10000000000 & 0.218 & Import (kW) & 30-04-2024 12:30 \\
10000000000 & 0.018 & Export (kW) & 30-04-2024 12:00 \\ 
10000000000 & 0.333 & Import (kW) & 30-04-2024 12:00 \\
~~~~... & ~~~~... & ~~~~... & ~~~~...\\
\bottomrule
    \end{tabular}
\end{table}

Data from smart meters installed in households and businesses was collected through a questionnaire. This  included questions about respondents' dwelling location (urban or rural) in order to determine which fixed-rate standing charge applies. Users  uploaded their HDF files as downloaded from the ESB Networks website and in return received an email containing estimates of their annual bill across all available tariff plans in the market, based on their own past electricity consumption.

The service was released publicly on \nth{31} January 2024 with ethics approval from the School of Computing ethics approval board. By \nth{15} June 2024, it had received 193 responses and uploads, representing 113 unique users. Some users uploaded their HDF files multiple times to obtain more recent feedback as tariffs changed over time. For these users, all data entries from the same  MPRN were combined by taking their earliest and latest end dates.  The collected data comprises 127 million data points across these 113 distinct users.

\subsection{Data Cleaning}

All smart meter data from users were trimmed to fall into the period from midnight on \nth{1} May 2023, to midnight on \nth{1} May 2024, as there was significant variation in the duration of uploaded HDF file records for each user, ranging from over two years to less than one month. 

HDF files downloaded from the ESB Networks website had occasional missing data which were sporadic and unpredictable, with the average deviation from the expected record count across all users being less than 0.43\%. The irregularities resulted from factors including network transmission problems, power outages, and other unidentified issues. For monthly records, if the missing data exceeded 10\% of the month's duration, the record was excluded from our dataset.

\subsection{Data Overview}

\label{subsec:gap}
After data cleaning, there were 107 users whose records covered at least one full calendar month. Among these, only 24 users had complete data for all 12 months because the smart meters for the others had been installed only within the previous 12 months. Such users are usually keen to use their smart meter data to find the best or cheapest tariff at the earliest possible opportunity, so are not willing to have to wait a full year after installation. Consequently, there is a real demand for predicting annual energy usage based on only partial data, as discussed later. Approximately 78\% of our users have HDF data spanning more than 6 months. The distribution of  data durations for these users is shown in Table~\ref{table:user_duration}.

\begin{table}[htb]
    \centering
    \caption{Number of users for different HDF data durations.
    \label{table:user_duration}}
    \begin{tabular}{l|l|l}
        \toprule
        \textbf{Duration} & \textbf{Number of Users} & \textbf{Percentage} \\ \midrule
        12 months & 24 & 22.4\% \\ 
        \verb|>| 9 months & 63 & 58.9\% \\ 
        \verb|>| 6 months & 83 & 77.6\% \\
        \verb|>| 2 months & 103& 96.3\% \\ 
        \verb|>| 1 month & 107 & 100\%\\ \bottomrule
    \end{tabular}
\end{table}

\begin{figure}[htb]
    \centering
    \includegraphics[width=0.8\linewidth]{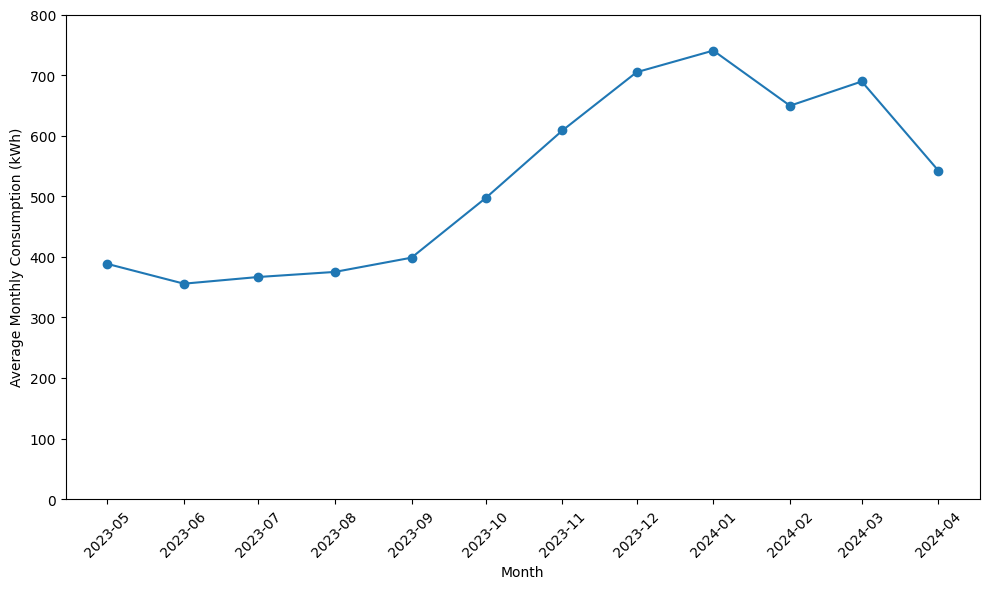}
    \caption{Average electricity consumption for each calendar month from 107 consumers.}
    \label{fig:Avg Monthly Consumption}
\end{figure}

Observing the average consumption of all our users over the  12 calendar months  reveals a  trend displayed in Figure~\ref{fig:Avg Monthly Consumption}. During the summer and warmer months from May to September, the consumption fluctuates around 377 kWh per month and peaks at approximately 740 kWh in January. The cold Winter   months from February to April still maintain a relatively high level of consumption, averaging around 627 kWh. The dip for February is because it has fewer days than other months.

Because of the varying durations in HDF records and the clear variation in usage based on seasonality, an annual consumption for those users who have missing HDF data as measured in months, cannot be accurately calculated as an average monthly consumption multiplied by 12. Therefore, in the following analysis of annual consumption, only the 24 users with complete data spanning the entire 12-month period are included in summary Table~\ref{table:annual consumption}. While the sample size is relatively small, the data presented, influenced by improving living standards,  challenges the accuracy of the CRU's recommended typical annual electricity consumption figure of 4,200 kWh, which was announced in 2017~\cite{cru4200} and on which all comparator websites for energy costs, including those regulated by the CRU such as bonkers.ie and switcher.ie, are based.

\begin{table}[htb]
    \centering
    \caption{Summary statistics for annual consumption among our 24 full-year users, in kWh}
    \label{table:annual consumption}
    \begin{tabular}{l|l|l|l|l|l}
        \toprule 
        Mean & 
        Min & 
        \nth{1} Quartile & \ 
        \nth{2} Quartile &  
        \nth{3} Quartile & 
        Max \\ \midrule
        7,125 & 1,704 & 4,370 & 6,503 & 8,711 & 22,639\\
        \bottomrule
    \end{tabular}
\end{table}

\begin{figure*}[ht]
    \centering
    \includegraphics[width=\linewidth]{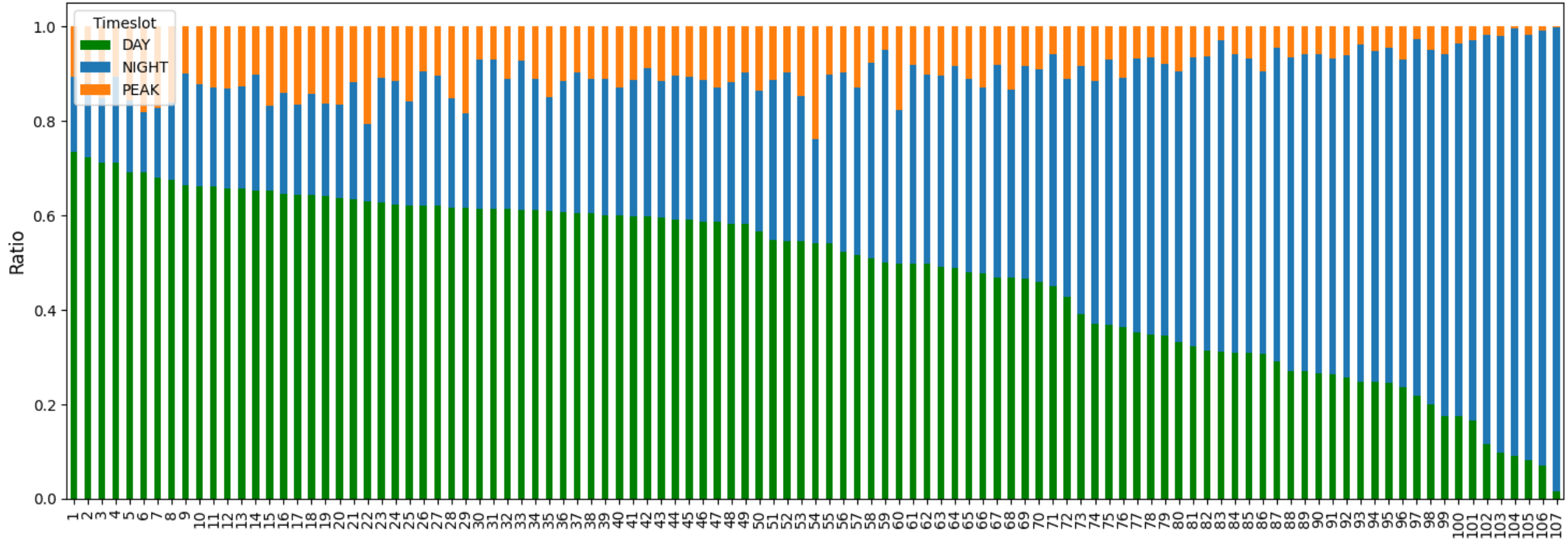}
    \caption{Consumption ratios for day, night and peak timeslots across users}
    \label{fig:Time of Use ratio across users}
\end{figure*}

\noindent  
An analysis of data from all 107 users provides insights into the broad spectrum of electricity consumption behaviours. Figure~\ref{fig:Time of Use ratio across users}  shows the ratios of electricity used across the 3 standard daily ToU timeslots for these 107 users. Here we see a notable portion (70) of these users display a trend of higher energy use during daytime, while 37 users show increased consumption patterns at night. 

\noindent
This analysis illustrates the  need for a more precise and personalised estimation of households' electricity consumption when selecting tariff plans, rather than relying on the CRU's stated annual domestic electricity consumption figure of 4,200 kWh.

\section{Methodology}

Users with complete HDF records spanning all 12 months had their past electricity usage calculated, and their estimated electricity bills for the subsequent 12 months were computed based on all available tariff plans. Here we assumed usage patterns would not change and that   consumption from the previous year would be the same as for the subsequent year.
For users with HDF records covering fewer than 12 months, a methodology was devised to estimate their electricity usage for the missing months with the collected dataset. This enabled the prediction of annual bills for these users, across different tariff plans.

\begin{figure}[!htb]
    \centering
    \includegraphics[width=0.6\linewidth]{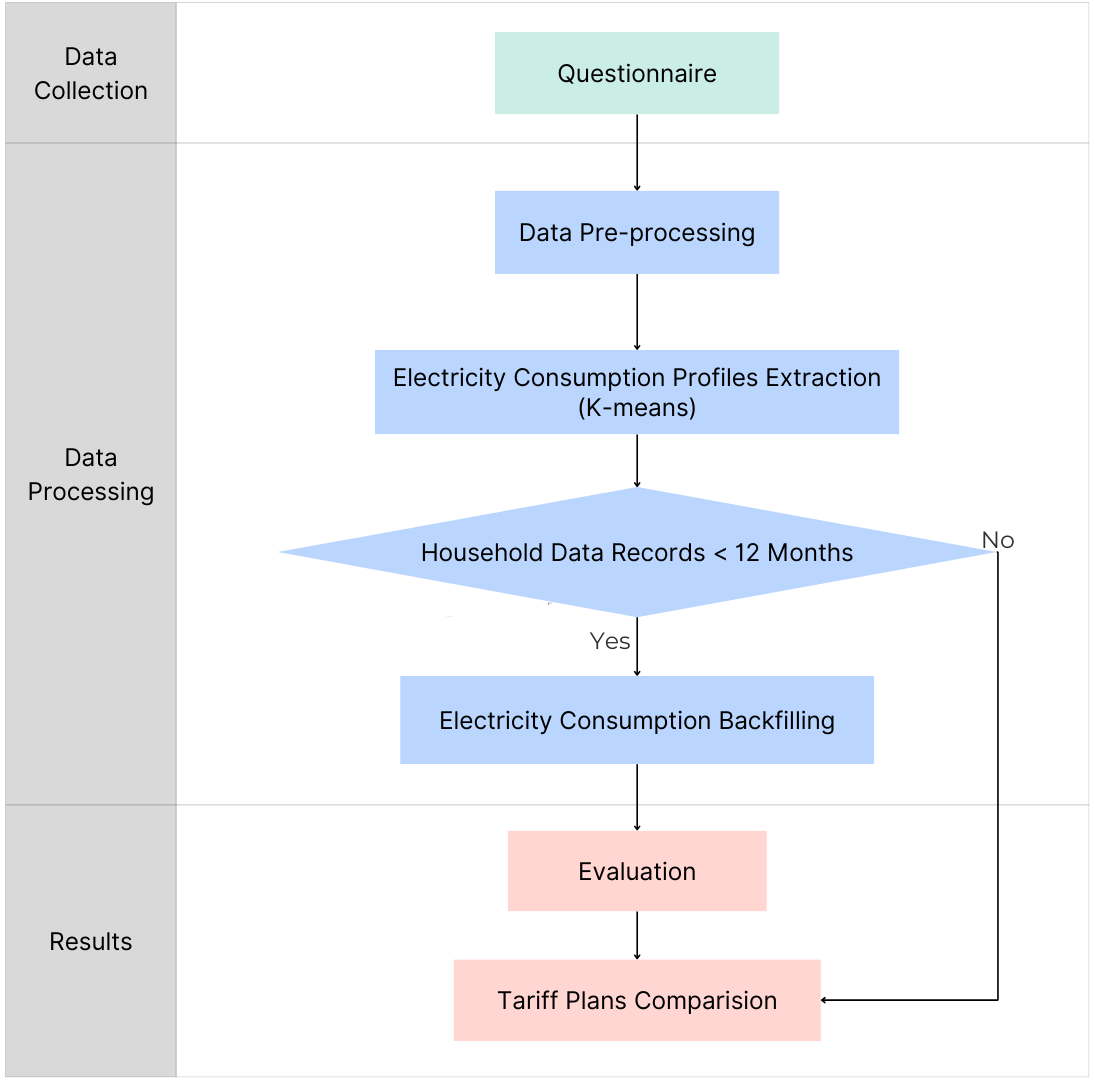}
    \caption{Workflow for estimating HDF data for missing months
    \label{fig:Workflow}}
\end{figure}

The overall methodology workflow is displayed in Figure~\ref{fig:Workflow}, including profile extraction, electricity usage prediction, evaluation and tariff plan comparison.

\subsection{Profile Extraction}

Our 107 users' HDF records were segmented into 12-month bins. For each month, the total usage was calculated for each of three different daily timeslots as used in ToU tariffs: day (8:00-17:00 \& 19:00-23:00), night (23:00-8:00), and peak (17:00-19:00). The average monthly usage for the three time periods were then calculated across the users. The monthly usage ratios for these periods were calculated relative to the total 12-month usage creating a series of data for each user, consisting of 36 ratios that sum to 1. Users were  clustered using K-means clustering based on the 36-dimension feature vectors.

As mentioned earlier the K-means~\cite{likas2003global} algorithm has  proven to be the most effective for processing smart meter data. It clusters data by partitioning it into k groups of equal variance, aiming to minimise  inertia criterion across clusters. 
%It begins by randomly initialising K cluster centroids and subsequently assigns each data point to the nearest centroid. The centroids are then recalculated based on the assigned points, and these steps are iteratively repeated until convergence is achieved. 
K-means is both simple and efficient, scaling effectively with a large number of samples and has been widely applied across various domains. In our study,  K-means used Euclidean distance between user profiles as the distance metric which can be expressed in terms of the Euclidean norm of the difference between p and q vectors, or users in our case:
\[distance=\|p-q\|_2=\sqrt{\sum_{i=1}^{n} (p_i - q_i)^2}\]

\begin{figure}[!htb]
    \centering
    \includegraphics[width=0.7\linewidth]{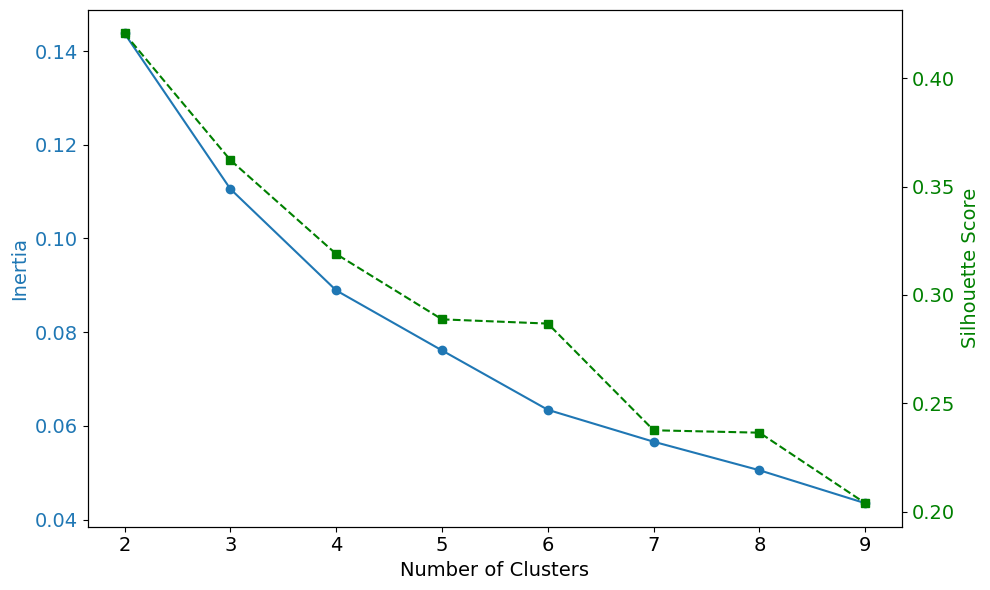}
    \caption{Elbow plot for K-means clustering on users.
    \label{fig:Inertia}}
\end{figure}

To determine the optimal number of user clusters, it is important to evaluate using both inertia and the Silhouette score, which are two widely used metrics for assessing clustering algorithms. Inertia~\cite{rykov2024inertia}, also referred to as the within-cluster sum of squares, quantifies the compactness of clusters and  decreases as the number of clusters increases.  Silhouette score~\cite{shahapure2020cluster} assesses how similar an object is to its cluster compared to other clusters, taking into account both cohesion (similarity within the same cluster) and separation (dissimilarity between different clusters).  Silhouette scores range from -1 to 1, with higher values indicating more well-defined clusters. Different numbers of clusters were tried in this investigation, and the results are shown in the elbow plot in Figure~\ref{fig:Inertia}. 

Balancing  Inertia and Silhouette scores is crucial and  Figure~\ref{fig:Inertia} suggests that using 5 or 6 clusters is the best choice. Despite the Silhouette scores being similar for 5 and 6 clusters, choosing fewer clusters is preferable, hence we chose 5 clusters. The centroid  for each cluster was computed and users with missing HDF data for certain months would be assigned to the appropriate cluster by adjusting cluster centroids  to match the lengths of available data for these users. This ensured accurate distance calculations to determine the closest cluster for each case of missing data.

\subsection{Back-filling Energy Consumption Data}

Users with at least 12 months of smart meter data have annual bills estimated for each tariff and the procedure for back-filling electricity consumption data for users with less than 12 months of data involves several steps and leverages similarities in electricity consumption patterns from their closest user cluster. This is shown in Figure~\ref{fig:Workflow}. In step 1, users were matched to the closest cluster profile based on only the number of months for which they have data.
In step 2 the day, night, and peak consumption ratios for the  missing months were determined based on that closest cluster profile.  Finally, in step 3, the estimated monthly energy consumptions for each time slot in each month was back-filled using the values from the cluster and normalised by the actual consumption figures for the months for which there is data.

\subsection{Evaluation of Clustering Accuracy}

We assessed the accuracy of our approach to back-filling HDF data by performing an evaluation  to assess clustering accuracy, verify if cluster assignments aligned with the closest cluster profile, and confirm the accuracy of back-filling for new users who submitted questionnaires after \nth{15} June 2024, and who have more than 12 months of data. 

As indicated by the data duration analysis  shown in Section~\ref{subsec:gap},  78\% of our users who uploaded their HDF files were missing up to six months of usage data from within the past year so we reconstructed historical usage data for periods ranging from one to six months. This involved taking users who had complete 12-month data and sequentially removing their oldest one to six months of data and then applying our back-filling process. Then the actual usage values could be compared with our estimated back-filled values for accuracy assessment.

To assess clustering accuracy, Symmetric Mean Absolute Percentage Error (SMAPE) was computed for each cluster profile during the day, night, and peak periods, as defined by the formula:
\[
\text{SMAPE} = \frac{100\%}{n} \sum_{i=1}^{n} \frac{|F_i - A_i|}{|A_i| + |F_i|}
\]
where \( F_i \) is the forecasted value, \( A_i \) is the actual value, and \( n \) is the number of forecasts.

SMAPE~\cite{HADJOUT2023120123} is intended to measure the relative difference between predicted and actual values, taking into account their magnitudes in a balanced way, where 0\% signifies a perfect prediction. SMAPE can effectively address the issue of having small or zero actual values in the denominator, which can lead to a very high error percentage. It offers a more balanced and robust measure as a result~\cite{HADJOUT2023120123}, compared to Mean Absolute Percentage Error (MAPE). The SMAPEs for the three standard ToU periods are then weighted based on their respective duration within the 24 hours —- 13 hours for day, 9 hours for night, and 2 hours for peak —- resulting in an overall error value. The clustering was deemed accurate if the assigned profile exhibited the best overall performance, as indicated by the lowest weighted SMAPE value.

When evaluating the feasibility of back-filling smart meter data for between one and six months, the objective was to confirm that this extended period could produce reliable predictions, as indicated by relatively low-weighted SMAPEs. This analysis would establish confidence in our back-filling process and ensure the reliability of future forecasts.

\section{Results}

\subsection{Creation of User Profiles}
\begin{figure}[h]
    \centering
    \includegraphics[width=0.5\linewidth]{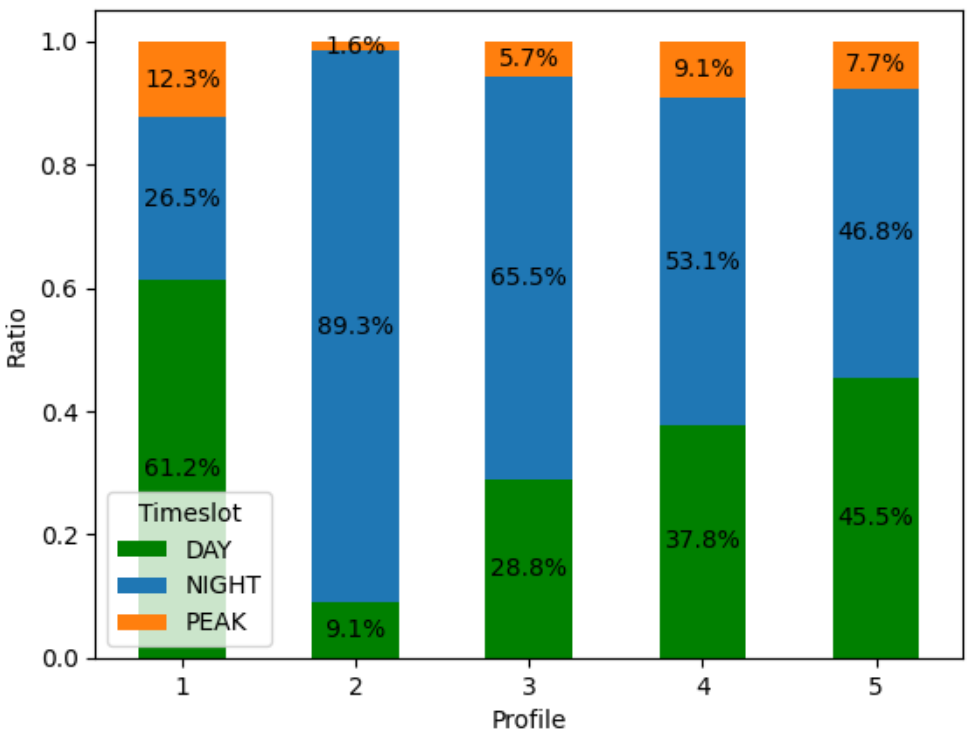}
    \caption{Consumption ratios for different ToU timeslots across 5 profiles/clusters.}
    \label{fig:Distribution of ToU across Profiles}
\end{figure}

\begin{figure*}[h]
    \centering
    \includegraphics[width=\linewidth]{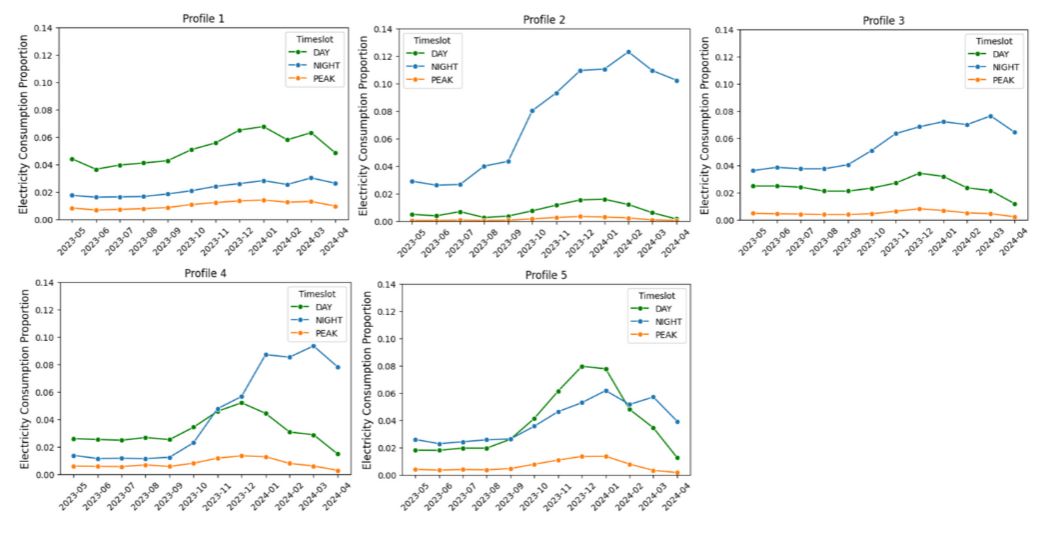}
    \caption{Five automatically determined electricity consumption usage profiles.}
    \label{fig:Profiles}
\end{figure*}

Five  consumption clusters were identified based on a K-means analysis of te 36 features of the 24 users' electricity usage patterns, all exhibiting a similar annual trend of increased usage in winter and decreased  in summer. We refer to the automatic clusters as profiles and across all profiles, users demonstrate relatively low electricity usage during peak times. However, there are  differences in their lifestyles and consumption behaviours between daytime and nighttime. The profile summaries  are shown in Figures~\ref{fig:Distribution of ToU across Profiles} and~\ref{fig:Profiles}.

As shown in Figure~\ref{fig:Profiles}, Profile 1, which we denote as the ``day profile" , is the most prevalent, comprising 55\% of households and predominantly consumes more electricity during daytime hours. In contrast, Profiles 2, 3, and 4, collectively representing 36\% of users, are categorised as ``night profiles" where households tend to use more energy during nighttime hours, especially during winter, possibly charging an EV or availing of cheaper energy to run a heatpump. Finally, Profile 5, which accounts for 9\% of households, displays a relatively balanced electricity consumption pattern between day and night but night usage outstrips day usage during the Winter months. These distinct consumption profiles highlight a range of lifestyle and consumption habits among different households.

\subsection{Evaluation of HDF Data Back-Filling}

The sample  users we used in this evaluation had submitted HDF files with full  coverage of the investigation period and as of July 2024, there were four such users (shown as A, B, C, and D in Table~\ref{tab:wsmape}) who met the criteria. 
Clustering results indicated that Users A and B were  assigned to Profile 1 for each of the six back-filling durations (back-filling for only the first month, then for the first two months, ..., and finally for the first six months). User C was clustered into Profile 3 for back-filling of their first to their first four months, and into Profile 4 for back-filling for five to six months. User D was assigned to Profile 5 across all six back-filling months.

The weighted SMAPE of estimated back-filling values vs. actual values are presented in Table~\ref{tab:wsmape}. 
For User A, when back-filled for one or two months, Profile 3 achieved the lowest SMAPE at approximately 15.3\%. However, when back-filled for more than two months, Profile 1 consistently achieved a low SMAPE, around 14.3\%, suggesting it provides the best overall accuracy. The pattern for User B is more pronounced, as the weighted SMAPEs for Profile 1 are consistently the smallest across all back-filling durations. For User C, the SMAPEs are lowest in Profile 3 when back-filling for up to five months whereas for the last back-fill duration, Profile 1 and Profile 3 have the same accuracy. The results from these sample of users validate the clustering outcomes. However, for User D, Profiles 1 and 4 show lower SMAPEs, which contrasts with our clustering which assigns User D to Profile 5. Further investigation revealed that User D's distances to the cluster centroids were very close across all profiles except Profile 2, suggesting that the user has a stochastic electricity usage pattern that does not align closely with any of our extracted profiles.

Overall, these findings validate the clustering outcomes, highlighting the reliability of our clustering methodology in capturing stable load patterns. However, accurately predicting energy consumption at the individual level may be less feasible for users exhibiting erratic load profiles and higher variability as supported by previous research~\cite{Kwac2013}.

\begin{table}[htb]
    \centering
    \caption{Averaged weighted SMAPE of 4 sample user profiles (A-D) across 6 back-filled months. Profile with lowest SMAPE for each test user for each back-fill month shown {\bf in bold}.}
    \label{tab:wsmape}
    \begin{tabular}{l|c|c|c|c|c|c}
        \toprule 
        Back-filling & \multirow{2}{*}{1} & \multirow{2}{*}{2} & \multirow{2}{*}{3} & \multirow{2}{*}{4} & \multirow{2}{*}{5} &\multirow{2}{*}{6}\\ 
        (months) &&&&&& \\
        \midrule
        Test user A\\
         Profile 1& 19.1\%&16.5\%& {\bf 14.9\%}&  {\bf 14.2\%}&  {\bf 14.3\%}& {\bf 13.6\%}\\
         Profile 2& 47.5\%&48.2\%&47.8\%&50.8\%&52.7\%&54.3\%\\
         Profile 3&  {\bf 15.3\%} &  {\bf 15.2\%}&16.7\%&18.4\%&19.2\%&21.3\%\\
         Profile 4& 20.3\%&22.5\%&25.6\%&29.2\%&29.9\%&30.4\%\\
         Profile 5& 15.4\%&17.5\%&19.9\%&23.6\%&21.7\%&20.4\%\\
         \midrule
         Test user B\\
         Profile 1& {\bf 7.9\%}&  {\bf 8.7\%}& {\bf 10.3\%}& {\bf 11.1\%}& {\bf 11.8\%}& {\bf 12.8\%}\\
         Profile 2&65.1\%&63.8\%&61.6\%&64.2\%&64.6\%&65.5\%\\
         Profile 3&36.6\%&36.8\%&38.9\%&40.8\%&42.3\%&44.2\%\\
         Profile 4&17.1\%&20.1\%&25.2\%&27.2\%&31.9\%&34.8\%\\
         Profile 5&39\%&37.7\%&38.4\%&39.4\%&38.6\%&37.7\%\\
         \midrule
         Test user C\\
         Profile 1& 25.7\%&23.3\%&22.6\%&23.4\%&24\%& {\bf 23.6\%}\\
         Profile 2& 55.9\%&60\%&60.8\%&63.4\%&63.7\%&62.4\%\\
         Profile 3&  {\bf 13.6\%}& {\bf 14.9\%}& {\bf 17.6\%}& {\bf 22.4\%}& {\bf 23.6\%}& {\bf 23.6\%}\\
         Profile 4& 29.3\%&33.9\%&37.9\%&41.9\%&44.2\%&42.4\%\\
         Profile 5& 29.3\%&33.8\%&36.2\%&40.1\%&39.8\%&35.1\%\\
         \midrule
         Test user D\\
         Profile 1& {\bf 8.3\%}&22.8\%&25\%&21.8\%&24.1\%&26.5\%\\
         Profile 2&52.6\%&55\%&55.8\%&60.4\%&57.7\%&55.4\%\\
         Profile 3&24.1\%&32\%&30.6\%&34.2\%&28.4\%&24.7\%\\
         Profile 4&24.2\%& {\bf 21.6\%}& {\bf 16.9\%}& {\bf 18.3\%}& {\bf 19.5\%}& {\bf 20.1\%}\\
         Profile 5&25.6\%&23.6\%&24.9\%&28.9\%&25.1\%&24.1\%\\

        \bottomrule
    \end{tabular}
\end{table}

For the stable users A, B, and C, the average weighted SMAPEs across six back-filling durations are approximately 15.4\%, 10.4\%, and 19.3\%, respectively. These weighted SMAPE values are less than 20\%, which is considered indicative of good forecasting performance~\cite{HADJOUT2023120123}. This confirms that back-filling for up to six months in our study is reliable 
despite the apparantly small amount of data used to create the clusters, while maintaining reasonable prediction accuracy and highlights our confidence in accurately estimating with six-month gaps in historical data.

Substantial savings are available in domestic energy bills when a user chooses the most economical  tariff for them, which may be ToU or fixed-rate depending on the amount and the characteristics of their use. It should be stressed that this analysis is based on usage patterns from profiles and that individual users may find that their most economical tariff may be either a ToU or a fixed-rate one. For that, we direct users to  
{\it URL blocked to preserve double-blind submission but available in final version}
%\url{http://tinyurl.com/4tjcwnea} 
where they can upload their own HDF data and where the  analysis on their on data is performed directly and the results sent to them.

\section{Conclusions}

In this study, smart meter usage (HDF) data from more than one hundred users were gathered and analysed to determine five typical consumption profiles. Due to  data collection time and acquisition channels, our sample size is relatively small compared to data gathered from national surveys. As the smart meter installation program is still ongoing in Ireland and elsewhere, many households  have had them installed within the last year, leading to a limited duration  usage data for such users. 

For users who do not yet have a full calendar year of their own historical HDF data, we developed and evaluated a method to estimate and back-fill their usage to allow an estimation of their full annual energy bill. This was done by categorising them into one of  five consumption profiles and using the usage patterns of the profile to complete the back-filling. The findings indicate that these estimates are reliable when predicting based on up to six months of missing data. Households in Ireland can choose from more than 60 tariffs from 10 suppliers but choosing the cheapest for each household can only be done using their smart meter data which captures each household's unique usage pattern. This work allows users to use a full year of their own smart meter data to find their best energy tariff in a way that accounts for seasonal factors, even if they have only 6 months of their own actual data.   %Additionally, the research reveals that ToU tariffs tend to consistently offer greater economic benefits for most users, especially those whose energy usage is primarily nighttime rather than daytime. 

In summary, this study provides  insights into an important area, namely estimating electricity consumption with a view to understanding the economic advantages of choosing the best tariff plan. It has the potential to benefit  stakeholders by deepening customers' understanding of their own energy use and helping them to choose the cheapest tariff for them.
Future work will explore the economic impact of different tariffs from different energy suppliers in terms of their effects on consumer's annual bills.

\begin{acknowledgments}

This work was partly-supported by Research Ireland under Grant Number: SFI/12/RC/2289\_P2, co-funded by the European Regional Development Fund.

\end{acknowledgments}

\bibliography{arxiv.bib}

\begin{thebibliography}{29}
\expandafter\ifx\csname natexlab\endcsname\relax\def\natexlab#1{#1}\fi
\providecommand{\url}[1]{\texttt{#1}}
\providecommand{\href}[2]{#2}
\providecommand{\path}[1]{#1}
\providecommand{\DOIprefix}{doi:}
\providecommand{\ArXivprefix}{arXiv:}
\providecommand{\URLprefix}{URL: }
\providecommand{\Pubmedprefix}{pmid:}
\providecommand{\doi}[1]{\href{http://dx.doi.org/#1}{\path{#1}}}
\providecommand{\Pubmed}[1]{\href{pmid:#1}{\path{#1}}}
\providecommand{\bibinfo}[2]{#2}
\ifx\xfnm\relax \def\xfnm[#1]{\unskip,\space#1}\fi
%Type = Misc
\bibitem[{{ESB Networks}(2023)}]{ESBN.2023}
\bibinfo{author}{{ESB Networks}}, \bibinfo{title}{{ESB Networks installs 1.5 million smart meters nationwide as part of the National Smart Metering Programme}}, \bibinfo{howpublished}{\url{https://tinyurl.com/3bdn9hus}}, \bibinfo{year}{2023}. \bibinfo{note}{Online; Accessed 21 November 2023}.
%Type = Misc
\bibitem[{Pope(2022)}]{smartcost}
\bibinfo{author}{C.~Pope}, \bibinfo{title}{{Will my smart meter cost me more and can I make it work for me?}}, \bibinfo{howpublished}{\url{https://www.irishtimes.com/ireland/2022/10/17/will-my-smart-meter-cost-me-more-and-can-i-make-it-work-for-me/}}, \bibinfo{year}{2022}. \bibinfo{note}{Online; Accessed 21 June 2024}.
%Type = Misc
\bibitem[{{ESB Networks}(2024)}]{esn}
\bibinfo{author}{{ESB Networks}}, \bibinfo{title}{{The National Smart Metering Programme}}, \bibinfo{howpublished}{\url{https://www.esbnetworks.ie/existing-connections/meters-and-readings/smart-meter-upgrade/background/}}, \bibinfo{year}{2024}. \bibinfo{note}{Online; Accessed 21 June 2024}.
%Type = Misc
\bibitem[{{Agency for the Cooperation of Energy Regulators}(2023)}]{ACER.2023.Report6}
\bibinfo{author}{{Agency for the Cooperation of Energy Regulators}}, \bibinfo{title}{{Demand response and other distributed energy resources: what barriers are holding them back?}}, \bibinfo{howpublished}{\url{https://www.acer.europa.eu/sites/default/files/documents/Publications/ACER_MMR_2023_Barriers_to_demand_response.pdf}}, \bibinfo{year}{2023}. \bibinfo{note}{Online; Accessed 19 December 2023}.
%Type = Article
\bibitem[{Bager and Mundaca(2017)}]{Bager}
\bibinfo{author}{S.~Bager}, \bibinfo{author}{L.~Mundaca},
\newblock \bibinfo{title}{{Making ‘Smart Meters’ smarter? Insights from a behavioural economics pilot field experiment in Copenhagen, Denmark}},
\newblock \bibinfo{journal}{Energy Research \& Social Science} \bibinfo{volume}{28} (\bibinfo{year}{2017}) \bibinfo{pages}{68--76}.
%Type = Article
\bibitem[{Belton and Lunn(2020)}]{Belton}
\bibinfo{author}{C.~A. Belton}, \bibinfo{author}{P.~D. Lunn},
\newblock \bibinfo{title}{{Smart choices? An experimental study of smart meters and time-of-use tariffs in Ireland}},
\newblock \bibinfo{journal}{Energy Policy} \bibinfo{volume}{140} (\bibinfo{year}{2020}).
%Type = Book
\bibitem[{Khan et~al.(2023)Khan, Mahela, Alhelou, and Padmanaban}]{marketderegulated}
\bibinfo{editor}{B.~Khan}, \bibinfo{editor}{O.~P. Mahela}, \bibinfo{editor}{H.~H. Alhelou}, \bibinfo{editor}{S.~Padmanaban} (Eds.), \bibinfo{title}{{Deregulated Electricity Market: The Smart Grid Perspective}}, \bibinfo{publisher}{Taylor \& Francis Group}, \bibinfo{year}{2023}.
%Type = Article
\bibitem[{Cassidy et~al.(2020)Cassidy, McLay, and Carmody}]{regulation}
\bibinfo{author}{E.~Cassidy}, \bibinfo{author}{P.~McLay}, \bibinfo{author}{W.~Carmody},
\newblock \bibinfo{title}{{Electricity Regulation 2021}},
\newblock \bibinfo{journal}{Lexology GTDT Series}  (\bibinfo{year}{2020}).
%Type = Misc
\bibitem[{{Electricity Supply Board}(2023)}]{ElectricitySupplyBoard}
\bibinfo{author}{{Electricity Supply Board}}, \bibinfo{title}{{Ratings Direct}}, \bibinfo{howpublished}{\url{https://cdn.esb.ie/media/docs/default-source/investor-relations-documents/s-p-credit-report-july-23.pdf}}, \bibinfo{year}{2023}. \bibinfo{note}{Online; Accessed 21 June 2024}.
%Type = Misc
\bibitem[{Pope(2011)}]{deregulated}
\bibinfo{author}{C.~Pope}, \bibinfo{title}{Electricity market to be deregulated}, \bibinfo{howpublished}{\url{https://www.irishtimes.com/business/energy-and-resources/electricity-market-to-be-deregulated-1.872715}}, \bibinfo{year}{2011}. \bibinfo{note}{Online; Accessed 21 June 2024}.
%Type = Misc
\bibitem[{{Commission for the Regulation of Utilities}(2024{\natexlab{a}})}]{CRU}
\bibinfo{author}{{Commission for the Regulation of Utilities}}, \bibinfo{title}{{What We Do}}, \bibinfo{howpublished}{\url{ https://www.cru.ie/about-us/what-we-do/}}, \bibinfo{year}{2024}{\natexlab{a}}. \bibinfo{note}{Online; Accessed 21 June 2024}.
%Type = Misc
\bibitem[{{Commission for the Regulation of Utilities}(2024{\natexlab{b}})}]{supplier}
\bibinfo{author}{{Commission for the Regulation of Utilities}}, \bibinfo{title}{{Energy Suppliers in Ireland}}, \bibinfo{howpublished}{\url{https://www.cru.ie/consumer-information/switch-supplier/energy-suppliers-in-ireland/}}, \bibinfo{year}{2024}{\natexlab{b}}. \bibinfo{note}{Online; Accessed 21 June 2024}.
%Type = Misc
\bibitem[{{Central Statistics Office}(2024)}]{census}
\bibinfo{author}{{Central Statistics Office}}, \bibinfo{title}{{Census of Population 2022 - Summary Results}}, \bibinfo{howpublished}{\url{ https://www.cso.ie/en/releasesandpublications/ep/p-cpsr/censusofpopulation2022-summaryresults/householdsizeandmaritalstatus/}}, \bibinfo{year}{2024}. \bibinfo{note}{Online; Accessed 21 June 2024}.
%Type = Article
\bibitem[{Andersen et~al.(2021)Andersen, Gunkel, Jacobsen, and Kitzing}]{ANDERSEN2021105341}
\bibinfo{author}{F.~Andersen}, \bibinfo{author}{P.~Gunkel}, \bibinfo{author}{H.~Jacobsen}, \bibinfo{author}{L.~Kitzing},
\newblock \bibinfo{title}{Residential electricity consumption and household characteristics: An econometric analysis of {D}anish smart-meter data},
\newblock \bibinfo{journal}{Energy Economics} \bibinfo{volume}{100} (\bibinfo{year}{2021}) \bibinfo{pages}{105341}. \DOIprefix\doi{10.1016/j.eneco.2021.105341}.
%Type = Article
\bibitem[{{Andreas Gunkel} et~al.(2023){Andreas Gunkel}, {Klinge Jacobsen}, Bergaentzlé, Scheller, and {Møller Andersen}}]{ANDREASGUNKEL2023108852}
\bibinfo{author}{P.~{Andreas Gunkel}}, \bibinfo{author}{H.~{Klinge Jacobsen}}, \bibinfo{author}{C.-M. Bergaentzlé}, \bibinfo{author}{F.~Scheller}, \bibinfo{author}{F.~{Møller Andersen}},
\newblock \bibinfo{title}{Variability in electricity consumption by category of consumer: The impact on electricity load profiles},
\newblock \bibinfo{journal}{International Journal of Electrical Power \& Energy Systems} \bibinfo{volume}{147} (\bibinfo{year}{2023}) \bibinfo{pages}{108852}. \DOIprefix\doi{10.1016/j.ijepes.2022.108852}.
%Type = Article
\bibitem[{Munkhammar et~al.(2015)Munkhammar, Bishop, Sarralde, Tian, and Choudhary}]{MUNKHAMMAR2015439}
\bibinfo{author}{J.~Munkhammar}, \bibinfo{author}{J.~D. Bishop}, \bibinfo{author}{J.~J. Sarralde}, \bibinfo{author}{W.~Tian}, \bibinfo{author}{R.~Choudhary},
\newblock \bibinfo{title}{{Household electricity use, electric vehicle home-charging and distributed photovoltaic power production in the city of Westminster}},
\newblock \bibinfo{journal}{Energy and Buildings} \bibinfo{volume}{86} (\bibinfo{year}{2015}) \bibinfo{pages}{439--448}. \DOIprefix\doi{10.1016/j.enbuild.2014.10.006}.
%Type = Article
\bibitem[{Wittenberg and Matthies(2016)}]{WITTENBERG2016199}
\bibinfo{author}{I.~Wittenberg}, \bibinfo{author}{E.~Matthies},
\newblock \bibinfo{title}{{Solar policy and practice in Germany: How do residential households with solar panels use electricity?}},
\newblock \bibinfo{journal}{Energy Research \& Social Science} \bibinfo{volume}{21} (\bibinfo{year}{2016}) \bibinfo{pages}{199--211}. \DOIprefix\doi{10.1016/j.erss.2016.07.008}.
%Type = Article
\bibitem[{Deng and Newton(2017)}]{DENG2017313}
\bibinfo{author}{G.~Deng}, \bibinfo{author}{P.~Newton},
\newblock \bibinfo{title}{{Assessing the impact of solar PV on domestic electricity consumption: Exploring the prospect of rebound effects}},
\newblock \bibinfo{journal}{Energy Policy} \bibinfo{volume}{110} (\bibinfo{year}{2017}) \bibinfo{pages}{313--324}. \DOIprefix\doi{10.1016/j.enpol.2017.08.035}.
%Type = Article
\bibitem[{Zhang et~al.(2021)Zhang, Wen, Li, Chen, Ye, Fu, and Livingood}]{ZHANG2021116452}
\bibinfo{author}{L.~Zhang}, \bibinfo{author}{J.~Wen}, \bibinfo{author}{Y.~Li}, \bibinfo{author}{J.~Chen}, \bibinfo{author}{Y.~Ye}, \bibinfo{author}{Y.~Fu}, \bibinfo{author}{W.~Livingood},
\newblock \bibinfo{title}{A review of machine learning in building load prediction},
\newblock \bibinfo{journal}{Applied Energy} \bibinfo{volume}{285} (\bibinfo{year}{2021}) \bibinfo{pages}{116452}. \DOIprefix\doi{10.1016/j.apenergy.2021.116452}.
%Type = Article
\bibitem[{Kang et~al.(2023)Kang, An, and Yan}]{KANG2023112753}
\bibinfo{author}{X.~Kang}, \bibinfo{author}{J.~An}, \bibinfo{author}{D.~Yan},
\newblock \bibinfo{title}{A systematic review of building electricity use profile models},
\newblock \bibinfo{journal}{Energy and Buildings} \bibinfo{volume}{281} (\bibinfo{year}{2023}) \bibinfo{pages}{112753}. \DOIprefix\doi{10.1016/j.enbuild.2022.112753}.
%Type = Article
\bibitem[{Michalakopoulos et~al.(2024)Michalakopoulos, Sarmas, Papias, Skaloumpakas, Marinakis, and Doukas}]{MICHALAKOPOULOS2024122943}
\bibinfo{author}{V.~Michalakopoulos}, \bibinfo{author}{E.~Sarmas}, \bibinfo{author}{I.~Papias}, \bibinfo{author}{P.~Skaloumpakas}, \bibinfo{author}{V.~Marinakis}, \bibinfo{author}{H.~Doukas},
\newblock \bibinfo{title}{A machine learning-based framework for clustering residential electricity load profiles to enhance demand response programs},
\newblock \bibinfo{journal}{Applied Energy} \bibinfo{volume}{361} (\bibinfo{year}{2024}) \bibinfo{pages}{122943}. \DOIprefix\doi{10.1016/j.apenergy.2024.122943}.
%Type = Article
\bibitem[{Rajabi et~al.(2020)Rajabi, Eskandari, Ghadi, Li, Zhang, and Siano}]{RAJABI2020109628}
\bibinfo{author}{A.~Rajabi}, \bibinfo{author}{M.~Eskandari}, \bibinfo{author}{M.~J. Ghadi}, \bibinfo{author}{L.~Li}, \bibinfo{author}{J.~Zhang}, \bibinfo{author}{P.~Siano},
\newblock \bibinfo{title}{A comparative study of clustering techniques for electrical load pattern segmentation},
\newblock \bibinfo{journal}{Renewable and Sustainable Energy Reviews} \bibinfo{volume}{120} (\bibinfo{year}{2020}) \bibinfo{pages}{109628}. \DOIprefix\doi{10.1016/j.rser.2019.109628}.
%Type = Article
\bibitem[{Czétány et~al.(2021)Czétány, Vámos, Horváth, Szalay, Mota-Babiloni, Deme-Bélafi, and Csoknyai}]{CZETANY2021111376}
\bibinfo{author}{L.~Czétány}, \bibinfo{author}{V.~Vámos}, \bibinfo{author}{M.~Horváth}, \bibinfo{author}{Z.~Szalay}, \bibinfo{author}{A.~Mota-Babiloni}, \bibinfo{author}{Z.~Deme-Bélafi}, \bibinfo{author}{T.~Csoknyai},
\newblock \bibinfo{title}{Development of electricity consumption profiles of residential buildings based on smart meter data clustering},
\newblock \bibinfo{journal}{Energy and Buildings} \bibinfo{volume}{252} (\bibinfo{year}{2021}) \bibinfo{pages}{111376}. \DOIprefix\doi{10.1016/j.enbuild.2021.111376}.
%Type = Misc
\bibitem[{{Commission for the Regulation of Utilities}(2024)}]{cru4200}
\bibinfo{author}{{Commission for the Regulation of Utilities}}, \bibinfo{title}{{Smart Meter Glossary}}, \bibinfo{howpublished}{\url{https://www.cru.ie/about-us/news/smart-meter-glossary/}}, \bibinfo{year}{2024}. \bibinfo{note}{Online; Accessed 21 June 2024}.
%Type = Article
\bibitem[{Likas et~al.(2003)Likas, Vlassis, and Verbeek}]{likas2003global}
\bibinfo{author}{A.~Likas}, \bibinfo{author}{N.~Vlassis}, \bibinfo{author}{J.~J. Verbeek},
\newblock \bibinfo{title}{The global k-means clustering algorithm},
\newblock \bibinfo{journal}{Pattern Recognition} \bibinfo{volume}{36} (\bibinfo{year}{2003}) \bibinfo{pages}{451--461}.
%Type = Article
\bibitem[{Rykov et~al.(2024)Rykov, De~Amorim, Makarenkov, and Mirkin}]{rykov2024inertia}
\bibinfo{author}{A.~Rykov}, \bibinfo{author}{R.~C. De~Amorim}, \bibinfo{author}{V.~Makarenkov}, \bibinfo{author}{B.~Mirkin},
\newblock \bibinfo{title}{{Inertia-based indices to determine the number of clusters in k-means: An experimental evaluation}} \bibinfo{volume}{12} (\bibinfo{year}{2024}) \bibinfo{pages}{11761--11773}. \DOIprefix\doi{10.1109/ACCESS.2024.3350791}.
%Type = Inproceedings
\bibitem[{Shahapure and Nicholas(2020)}]{shahapure2020cluster}
\bibinfo{author}{K.~R. Shahapure}, \bibinfo{author}{C.~Nicholas},
\newblock \bibinfo{title}{{Cluster quality analysis using silhouette score}},
\newblock in: \bibinfo{booktitle}{2020 IEEE 7th International Conference on Data Science and Advanced Analytics (DSAA)}, \bibinfo{organization}{IEEE}, \bibinfo{year}{2020}, pp. \bibinfo{pages}{747--748}.
%Type = Article
\bibitem[{Hadjout et~al.(2023)Hadjout, Sebaa, Torres, and Martínez-Álvarez}]{HADJOUT2023120123}
\bibinfo{author}{D.~Hadjout}, \bibinfo{author}{A.~Sebaa}, \bibinfo{author}{J.~F. Torres}, \bibinfo{author}{F.~Martínez-Álvarez},
\newblock \bibinfo{title}{Electricity consumption forecasting with outliers handling based on clustering and deep learning with application to the {A}lgerian market},
\newblock \bibinfo{journal}{Expert Systems with Applications} \bibinfo{volume}{227} (\bibinfo{year}{2023}) \bibinfo{pages}{120123}. \DOIprefix\doi{10.1016/j.eswa.2023.120123}.
%Type = Inproceedings
\bibitem[{Kwac et~al.(2013)Kwac, Tan, Sintov, Flora, and Rajagopal}]{Kwac2013}
\bibinfo{author}{J.~Kwac}, \bibinfo{author}{C.-W. Tan}, \bibinfo{author}{N.~Sintov}, \bibinfo{author}{J.~Flora}, \bibinfo{author}{R.~Rajagopal},
\newblock \bibinfo{title}{Utility customer segmentation based on smart meter data: Empirical study},
\newblock in: \bibinfo{booktitle}{2013 IEEE International Conference on Smart Grid Communications (SmartGridComm)}, \bibinfo{year}{2013}, pp. \bibinfo{pages}{720--725}. \DOIprefix\doi{10.1109/SmartGridComm.2013.6688044}.

\end{thebibliography}

\end{document}